# Resistance Management for Cancer: Lessons from Farmers

Sareh Seyedi[1,2,3], Valerie K. Harris[1,2], Stefania E. Kapsetaki[1,2], Shrinath Narayanan[2,4], Daniel Saha[1,2,3], Zachary Compton[1,2,3], Rezvan Yousefi[1,2,5], Alexander May[6], Efe Fakir[7], Amy M. Boddy[1,8,9], Marco Gerlinger[10,11], Christina Wu[12], Lida Mina[13], Silvie Huijben[3,14], Dawn H. Gouge[15], Luis Cisneros[1,2,3], Peter C. Ellsworth[15], Carlo C. Maley[1,2,3,14]

**Affiliations**
[1] Arizona Cancer Evolution Center, Arizona State University, Tempe, AZ, USA
[2] Center for Biocomputing, Security and Society, Biodesign Institute, Arizona State University, Tempe, AZ, USA
[3] School of Life Sciences, Arizona State University, Tempe, AZ, USA
[4] Department of Ecology and Evolution, University of Lausanne, Switzerland
[5] The Polytechnic School, Ira A. Fulton Schools of Engineering, Arizona State University, USA
[6] Research Casting International, Quinte West, ON, Canada
[7] Istanbul University Cerrahpasa School of Medicine, Turkey
[8] Exotic Species Cancer Research Alliance, North Carolina State University, Raleigh, NC, 27607 USA
[9] Department of Anthropology, University of California Santa Barbara, CA, USA
[10] Translational Oncogenomics Laboratory, Centre for Evolution and Cancer, The Institute of Cancer Research, London, SW3 6JB, UK
[11] Gastrointestinal Cancer Unit, The Royal Marsden Hospital, London, SW3 6JJ, UK
[12] Division of Hematology/Medical Oncology, Department of Medicine, Mayo Clinic, Arizona, USA
[13] Mayo Clinic, Phoenix, Arizona, USA
[14] Center for Evolution and Medicine, Arizona State University, USA
[15] Department of Entomology, University of Arizona, USA

**Abstract**

One of the main reasons we have not been able to cure cancers is that drugs select for drug-resistant cancer cells. Pest managers face similar challenges with pesticides selecting for pesticide-resistant organisms. Lessons in pest management have led to four heuristics that can be translated to controlling cancers: (1) limit use (of chemical controls / modes of action to the lowest practical levels); (2) diversify use (of modes of action largely through rotations of chemical controls); (3) partition chemistry (modes of action through space and time, which in effect is a refuge management strategy); and (4) include non-chemical methods. These principles are general to all cancers and cancer drugs, and thus should be employed to improve oncology. We review the parallel difficulties in controlling the evolution of drug resistance in pests and cancer cells, and describe the results of single- and multi-drug strategies in agriculture and oncology. We dissect the methods that pest managers use to prevent the evolution of pesticide resistance, showing how integrated pest management inspired the development of adaptive therapy in oncology to stabilize tumor size, and increase progression-free survival and patients' quality of life. Finally, we demonstrate these principles in a proposal for clinical trials in colorectal cancer.

**Glossary**

Integrated Pest Management (IPM): A comprehensive approach to controlling pests by emphasizing multifaceted, system-based strategies from multiple disciplines.

Mode of Action (MoA): The biochemical mechanism by which a drug achieves its effect.

Lethal Time 50 (LT50): The median time until the death of an organism after exposure to a toxin or stressful condition.

Synthetic pyrethroid (SP): A pesticide derived from naturally occurring pyrethrins, taken from the pyrethrum of dried *Chrysanthemum* flowers.

Chimeric antigen receptor T cells (CAR-T cells): T cells genetically engineered to produce an artificial T cell receptor for use in immunotherapy.

PD-L1/PD-1: PD-1 is expressed on T cells and PD-L1 on cancer cells and other cells in a tumor. The PD-1 and PD-L1 is a receptor-ligand system the result of their attachment is the blockade of anti-tumor immune responses.

Circulating tumor DNA (ctDNA): Tumor-derived, free-floating fragmented DNA in the bloodstream that is not inside cells.

Single-photon emission computed tomography (SPECT/CT): A tomographic imaging technique in nuclear medicine that uses gamma rays.

Standardized uptake value (SUV): A semiquantitative measure in clinical FDG-PET/CT oncology imaging that shows the degree of metabolism in a tissue.

Patient-derived xenograft (PDX): Tissue or cells from a patient's tumor that have been implanted into another organism, usually an immunodeficient or humanized mouse.

Autochthonous tumors: Tumors that are spontaneous or induced by chemical, viral, or physical factors in wild-type hosts, rather than arising in a genetically engineered host or transplanted from another host.

Pesticide resistance: A heritable, statistically defined decrease in sensitivity to a control agent in a pest population relative to the response of susceptible populations never previously exposed.

Cross-resistance: Single genetic factor confers resistance to multiple pesticides with the same or a related MOA.

Multiple resistance: Resistance to multiple drugs.

Action threshold: The number of pests or level of pest-induced damage that needs to be exceeded before action is taken.

Multidrug therapy: Using a mixture of two or more drugs with a different modes of mechanism of action.

Management connotes that some level of pest presence is tolerable. Control implies the actions taken to reduce or eliminate pests.

Intratumor heterogeneity (ITH): tumor cell populations within the same tumor have different phenotypic and molecular profiles.

**Introduction**

Oncologists and pest managers have similar problems in that both cancer cells and pest populations are composed of large numbers of genetically diverse organisms spread over heterogeneous (micro)environments[1–3]. When pest managers apply a pesticide to kill pests and oncologists apply a drug to cancer cells, they kill the sensitive pests and cancer cells but select for variants that are resistant to the drug[4,5]. In fact, most cancer deaths are caused by the evolution of therapeutic resistance[6–8]. Both fields also have the challenge of limiting collateral damage and other negative side effects due to the toxicities of their drugs[9–11].

Pesticides have off-target effects on other organisms in the environment that include those beneficial to the management system like natural enemies and pollinators similar to the toxic systemic effects of chemotherapeutic modalities/drugs that impact non-target tissues and cells important to the health of the patient[12–14].

To date, almost all cancer therapies select for therapeutic resistance[8]. Failure to address the evolutionary dynamics of cancer will result in the sustained high mortality of the metastatic disease and future cancer therapies will inevitably suffer the same fate. Decades of experimentation in pest management have led to a series of insights and effective heuristics for controlling pests called Integrated Pest Management (IPM) that, for the most part, have never been tried in oncology. What can oncologists learn from pest managers and how might we translate those insights to the clinic?

IPM is a comprehensive approach to pest management emphasizing multifaceted, system-based strategies involving multiple disciplines[15]. IPM was initially introduced as "integrated control" in 1959 by Stern et al., who synthesized aspects of biological, spatial, and chemical control tactics to reduce pest populations below levels of economic concern while limiting the development of resistance within agricultural pest communities[16]. Complete eradication of pests with pesticides is usually impossible because pest populations are large, and genetically diverse, and new pests can readily migrate into fields from neighboring areas[17,18]. Given these circumstances, pest managers have come to focus on long-term management methods rather than eradication. IPM has achieved successful, long-term therapeutic resistance management by diversifying control mechanisms and avoiding over-reliance on single-action pest control methods. By viewing resistance management strategies as a dynamic process rather than the application of a fixed protocol, IPM has been able to address and harness the power of genetic adaptation to control therapeutic resistance[11].

Translated to oncology, an analogous approach would be to control cancers so that patients could live with the disease, but not die from it. While this might seem like a radical change in how we treat cancer, it is already the de facto goal in many of the treatments of metastatic cancers in which cures are extremely unlikely[19,20]. Overall, IPM aims to keep pest populations below an economically acceptable threshold while limiting the use of pesticides in order to preserve the environment and human health[11]. Indeed, low levels (sub-economic) of pests are tolerated in part because they are fodder for the natural enemies and help maintain their densities in fields where biological control is needed either in the control of same pests or other pest species that arrive[21].

Here we briefly review the history of the emergence of acquired therapeutic resistance and early attempts to address it by using multiple drugs in both pest management and oncology. We then review the principles of integrated pest management[22] and discuss how they might be translated to oncology. Finally, we review early promising results in the application of those principles to oncology and provide an example of a potential future clinical trial in colorectal cancer, based on IPM.

**Single-drug therapies tend to fail**

As early as 1914, pest managers observed that single-agent pesticide treatments did not work long-term because they were selected for pesticide resistance[13]. While the 1960s-70s saw a dramatic upswing in pesticide development, even with the introduction of new insecticides such as organophosphates, carbamates, synthetic pyrethroids, and other potent agrochemicals, pest populations still evolved mechanisms of resistance [23–25]. In Australia, the use of DDT led to the evolution of DDT resistance in the cotton bollworm moth (*Helicoverpa armigera*). Synthetic pyrethroids (SP) were highly effective in the late 1970s and in 1978, pest managers were able to achieve 76% control of *H. armigera* to protect the cotton industry[26]. However, by 1983 that had dropped to only 10%[26]. Moreover, insects developing resistance even to cultural controls including things like crop rotation also mentioned that two species of *Diabrotica* have also evolved resistance to crop rotation[27,28].

Acquired therapeutic resistance was also documented in the first clinical reports of chemotherapy for cancer[29,30]. Single-drug therapies, such as nitrogen mustard and aminopterin, were used in the treatment of tumors in the late 1940's[29,30]; however, acquired resistance to treatment was noted against both drugs[29,30]. Resistance also evolved in response to radiotherapy, which helped drive interest in developing chemotherapies as a main method to treat cancer[29]. Targeted therapies, advanced biologic drugs, and even immune-checkpoint-blockade therapies also select for therapeutic resistance[8,31].

**Multidrug therapies are better than single drugs but still tend to fail**

In theory, if resistance to a single drug is rare in a population of pests or cancer cells and the mechanism of resistance is different for each drug, resistance to multiple independent drugs should be much more rare than resistance to a single drug. Experience has shown that this is not the case, however. Even though several studies have shown combination therapy is better than monotherapy in both pest management and cancer therapy, combination therapy still selects for multi-drug resistance and ultimately fails[32–39]. This is in part due to cross-resistance between drugs and mechanisms of multi-drug resistance such as the up-regulation of efflux pumps that can provide resistance to drugs with independent modes of action (MoAs)[26,37,40,41]. Combining drugs has therefore not solved the problem of acquired therapeutic resistance in most cases.

*Evidence in pest management*

In 1989, Tabashnik provided a detailed review of the failures of multiple drug usage in pest management[5]. The combinatorial use of insecticides in several studies showed that this method could not significantly prevent resistance[32–36]. There are important concerns regarding the combinatorial use of drugs, including toxicity for the natural predators of the pests, development of resistance in secondary pests, and increased selective pressure for cross-resistance[5]. mixtures of insecticides outperform sequential therapies in terms of deferring resistance[42,43]. However "sequential" is defined as using product A until it no longer works (because of resistance) and then switching to product B. A study conducted by Djouaka et al. on populations of *Anopheles funestus* mosquitoes in southwest Nigeria[44] revealed the presence of multi-drug resistance to a combination of different insecticides such as DTT, Dieldrin, and Permethrin. The mosquitoes were able to withstand lethal doses of these drugs[44].

Burden et al. tested 4 different approaches to prevent the development of resistance in the German cockroach when exposed to the insecticides chlordane and malathion[32]. One colony was exposed to chlordane, another colony to malathion, another colony was exposed to both compounds, and a fourth colony was exposed to each drug sequentially, switching drugs each generation. They found that resistance developed in both colonies that used insecticides in combination (either using in tandem or sequentially) nearly as fast as in the colony that was exposed to just chlordane[32].

A study by Ozaki in 1983 tested the combination of pesticides with potentially synergistic effects and different modes of action[33,45]. They tested malathion and carbaryl on small brown planthoppers (*Laodelphax striatellus*). They applied the insecticides alternatively every other generation. While alternating pesticides with different modes of action delayed the evolution of resistance compared to single pesticide therapy, multi-pesticide resistance developed after four generations and an eventual 11-fold increase of resistance by the 12th generation[45]. They also found that combinations of two or more insecticides that have synergistic effects could cause delays in the development of resistance, but multi-insecticide resistance still evolved over time.

There is a study conducted by MacDonald et al. in 1983, three different treatment regimes on the development of permethrin and dichlorvos resistance for house fly control: Single-use of each insecticide and use of both of them in rotation[35]. Overall, resistance most rapidly emerged in single-use treatment groups. The alternating use of permethrin and dichlorvos reduced resistance evolution and delayed resistance in the field, which is due to the less selective pressure with any of the components[35].

Another study by Pimentel and Burgess showed that using antagonistic effects among different chemical, physical, and biological selective pressures could reduce the development of resistance to insecticides and cross-resistance between them[34]. They tested 5 insecticides on house flies (*Musca domestica*) using different strategies: (1) Single versus multiple insecticides, (2) different order and rotations of insecticides, and (3) combinations of insecticides with biological controls (e.g., predators) and physical factors (e.g., insect traps)[34]. They found that the application of insecticides in rotation and combinations of insecticides was better at suppressing the development of resistance than the single-use of insecticide sequentially[34].

They have used these five drugs in diverse sequences and rotations (changed every generation) and found there was less cross-resistance than the single use of an insecticide sequentially. However, after 30 generations, some insecticides showed developing more resistance when they used the rotation and mixture method compared to single-use of insecticide sequentially[34].

Georghiou et al. (1983) studied three insecticides (temephos, permethrin, and propoxur) on the *Culex quinquefasciatus* strain of mosquitos which had genes encoding for resistance to each of these insecticides in the laboratory[36]. Both temephos and propoxur are designed to inhibit the normal breakdown of acetylcholine, so they have a similar mode of action. These three insecticides were used singly and in various combinatorial methods. Overall, resistance evolved rapidly when a single insecticide was used and resistance was suppressed by a combination of insecticides[36]. However, resistance still evolved even under a combination of some modes of action (e.g., when propoxur was in the combination)[36]. Moreover, there was a rapid reduction of resistance in a rotation strategy for some modes of actions (e.g., temephos and permethrin)[36].

Therefore, it should be noted that besides the strategy of the treatment, the synergistic effect of the combined drugs is also important. While combining insecticides remains a theoretical tactic for delaying resistance in the case of pest management, mixtures that meet the optimization criteria of models remain elusive in practical application[43].

*Evidence in cancer therapy*

The standard of care in the field of cancer treatment is to combine drugs[41]. What is less widely acknowledged is that these multidrug therapies also tend to fail, particularly in late stage cancers, leading to multidrug resistance[37]. Meta-analyses of lung[41,46,47], breast[48], gastric[48,49], pancreatic[50–52], and ovarian cancers[53,54] show modest survival benefits of combining drugs with increased toxicity but rarely cures from complex combination therapies.

Another meta-analysis of 35 randomized phase III trials in NSCLC compared single drug therapy versus 2 or 3 drugs[41]. This study showed that there was a significant increase in both tumor response (OR, 0.42; 95% CI, 0.37-0.47; P<0.001) and 1-year survival (OR, 0.80; 95% CI, 0.70-0.91; P<0.001) using doublet regimes compared to single drug therapy. In addition, they found an increase in the tumor response rate when 3 drugs were used in combination (OR, 0.66; 95% CI, 0.58-0.75; P<0.001). However, there was no increase in 1-year survival when the third drug was added (OR, 1.01; 95% CI, 0.85-1.21; P= 0.88). Overall, greater toxicity was also observed in combination chemotherapy compared to single-agent chemotherapy[41].

Another study on randomized phase III trials compared paclitaxel alone versus paclitaxel and carboplatin in advanced NSCLC and found no statistically significant improvement in overall survival in the combined therapy[46]. The 1-year survival was 32% and 37% for single and combination therapy, respectively with a hazard ratio of 0.91 (95% CI, 0.77 to 1.17; p= 0.25)[46].

A more recent meta-analysis on patients with advanced NSCLC from phase II and III s clinical trials comparing combination chemotherapy as second-line therapy with single-agent chemotherapy in the treatment of patients found no significant difference in overall survival (OS) between the combination and single therapy ($P = 0.32$). Patients under combination therapy showed more grade 3 (28%) to 4 (22%) nonhematologic toxicity ($P = 0.034$) and substantially grade 3 (41%) to 4 (25%) hematologic ($P = 0.0001$)[55].

In breast cancer, a meta-analysis carried out on 43 randomized controlled trials included 9,742 patients who received chemotherapy in combination or single-agent as their first-line treatment[48]. They found that combination therapy improved overall survival and time to progression significantly (HR 0.88, 95% CI 0.83- 0.93, p<0.00001) compared to single-agent therapy[48]. However, women who received combination therapy faced serious side effects such as vomiting and nausea due to the toxicity result of treatment[48].

A meta-analysis of first-line chemotherapy in advanced gastric cancer on randomized phase II and III trials found that combination therapy (2 or 3 drugs) improved survival (HR=0.83, 95% CI:0.74-0.93) compared to single therapy. However, toxicities related to the treatment were higher in combination strategies [56]. A study comparing the combination of 3 drugs (docetaxel, cisplatin, and fluorouracil) vs 2 drugs (cisplatin and fluorouracil) in phase III as first-line therapy of advanced gastric cancer found that overall survival (P= 0.02) and time to progression (p< 0.001) were longer for 3 drugs compared to 2 drugs. In addition, the 2-year survival rate on 3 drugs was 18% versus 9% for two drugs. However, adding the third drug led to higher toxicity[57] and with only an 18% 2-year survival, clearly did not solve the problem of therapeutic resistance.

Meta-analyses on phase II and III s clinical trials in pancreatic cancer have shown an advantage in using multidrug therapies, with improvements in tumor response and patient survival but increased toxicities compared to single drug therapies and rarely any cures. Heinemann et al. (2008) analyzed advanced or metastatic pancreatic cancer patients and found that there was an improvement in overall survival using a combination regimen containing gemcitabine along with another cytotoxic drug versus gemcitabine alone (Hazard ratio:0.91, 95% CI:0.85-0.97, P = 0.004)[51].

Another study by Moore et al. on 569 patients of randomized phase III trials with advanced pancreatic cancer showed both overall survival (HR= 0.82, P=0.083, 95%CI 0.69 to 0.99) and 1-year survival (23% vs 17%, P= 0.023) improvements using a combination of gemcitabine with erlotinib compared to using gemcitabine alone[58]. Progression free survival was also significantly better in this combination compared to single agent therapy (HR= 0.77, 95%CI 0.64 to 0.92; P= 0.004). However, there were more toxicities with combination treatment and few cures.

Pusceddu, et al. (2019) conducted a meta-analysis study on randomized phase III trials on metastatic pancreatic cancer (mPC) patients using a 2-drug combination of gemcitabine with nab-paclitaxel (GEM-NAB) versus a 3-drug combination, FOLFIRINOX (5-fluorouracil, oxaliplatin, and irinotecan)[59]. In terms of median overall survival, the FOLFIRINOX was better compared to GEM-NAB (11.1 vs. 8.5; p=0.03, 95% CI 0.08-2.22). However, there was no difference in time to progression[59].

A meta-analysis of four randomized II and III trials with a total of 1,300 ovarian cancer patients comparing single-agent platinum chemotherapy versus platinum combined with another chemotherapy (e.g., Carboplatin with Gemcitabine), in women with relapsed platinum-sensitive ovarian cancer, found an improvement in overall survival for combination platinum chemotherapy (HR = 0.80; 95%CI: 0.64-1.00, P = 0.05)[53]. Two of the four trials included quality of life data, which showed there was no adverse effect on the quality of life between the single and combination-platinum treatments[53]. The 3-year overall survival of using carboplatin alone was 29% compared to 42% of using carboplatin with epidoxorubicin (odds ratio = 0.8; 95% CI, 0.6–1.2). However, platinum combination therapy led to higher toxicity[53].

In virtually all of these meta-analyses, investigators found evidence of increased efficacy for multiple drugs over single drugs, often at the cost of increased toxicity and treatment-related deaths.

However, we must face the fact that while multidrug therapy may extend life by a matter of months, it rarely achieves a cure in late-stage disease, and there are diminishing returns from adding more drugs due to increases in toxicity and the evolution of multidrug resistance. In other words, multidrug therapy does not solve the problems of cancer or pest infestations.

**Modern principles of resistance management**

Through decades of experimentation and observation, pest managers have learned to focus on management, rather than eradication of pests[60]. They have developed a set of principles for maintaining long-term management of pest populations[11,22] (Figure 1):
- Prevention and suppression - adjust crops and the habitat to suppress proliferation of the pests and favor the crops.
- Monitor continuously, including systems for early warnings.
- Decision making - identify acceptable thresholds of infestation, below which there will be no intervention and change interventions in response to changes in pest levels.
- Prefer non-chemical control methods, including mechanical, biological and habitat controls. But when chemicals have to be used,
- Use pesticides as specific as possible to the type of pest, minimizing off-target effects, including pesticides that modify pest behavior rather than killing them.
- Use the lowest effective dose possible.
- Choose pesticides so as to reduce cross-resistance, and separate their application in both space and time.
- Evaluate success based on long-term management.
- Forecasting of pest growth and response to interventions, informs most of the other principles.

Pest managers are very clear that no single intervention is sufficient to control pests, and that usually the use of some chemicals is necessary[11]. However, combining these principles, and combining chemical controls with other forms of controls can lead to the effective and long term management of pests[11], hence the emphasis on both *integration* and *management* in the term Integrated Pest Management.

[FIGURE 1]

**Figure 1: Lessons from Integrated Pest Management to oncology.** Best practices in resistance management that we can apply in cancer treatment.

*Prevention and suppression*

Cancer biologists, like pest managers, have noted that prevention should be a key part of the strategy for dealing with cancer[1,3,9,12,13,61]. It is much easier to prevent an infestation or cancer than it is to treat it, and as such, full commitment to support efforts and resources devoted to cancer prevention is critically necessary. Despite this knowledge, there is currently and likely will remain, a pressing need to treat both pest infestations and cancers that have not been or are not able to be prevented. To that end, there are a number of strategies that may be employed to suppress pest populations. These include making the crops more robust to infestation through rotating crops, increasing the genetic diversity of crops (so that it is difficult for pests to specialize on one type of host), and changing the habitat of the field to suppress the pests and/or favor the crops[11,15,16] . Because stressed plants are more susceptible to pest infestation, pest populations can be suppressed by increasing crop health and reducing stress on plants through the use of fertilizers, timely irrigation and even plant growth regulators[62–64].

Translating these ideas to oncology might involve modifying the microenvironment of a tumor to favor healthy cells and disfavor cancer cells[65], which we discuss below under habitat controls. Other strategies, such as improving the overall health of patients through diet and exercise, and other lifestyle factors, could be helpful in making the tissue microenvironments less conducive for cancer[66–68].

*Monitor continuously and change interventions*

Traditionally, oncologists do not monitor tumor burden during a therapy protocol and only check weeks to months later to determine whether the therapy is effective. For example, in a clinical study that was conducted on patients with rectal adenocarcinoma, their response to chemoradiation was determined within 2-4 weeks after the treatment completion[69]. In cases where a solid tumor is monitored during therapy, response evaluation criteria in solid tumors (RECIST) guidelines are to evaluate it every 6-8 weeks[70], though every 12 weeks is common[71]. This prolonged period between treatment and monitoring makes it difficult to get an early warning that a tumor's response to therapy is slowing, indicating the emergence of resistance and the need to change the intervention. More frequent monitoring can reveal an initial response followed by regrowth that might appear as a failure to respond with less frequent monitoring. The kinetics of cancers and their responses will vary from cancer to cancer, and even within the same cancer type, from patient to patient[39]. The frequency of measuring tumor response necessary to manage resistance is an open question, but it is likely to be more frequent than current practice. For these purposes, liquid biopsies and ctDNA are particularly attractive as minimally invasive methods for monitoring tumor burden, which might additionally reveal the evolution and expansion of resistant clones[72–76].

*Decision making*

It is common in IPM to avoid using pesticides as long as damage to the crops is below acceptable levels[11]. In fact, such thresholds are central to pest management. Avoiding the use of a drug is obviously one of the best ways to avoid selecting for resistance to that drug. Translated to oncology, this would suggest watchful waiting as long as the tumor is small enough and there are no signs of significant damage to the patient. It is an open question how to measure the damage or danger posed by a particular tumor, but the principle is to establish some threshold, below which we do not treat. These thresholds may be based on multiple objectives such as quality of life, risk of progression, performance status, and other measures of health. It is important to keep in mind that we are assuming the presence of resistance, so withholding therapy does not imply missing the opportunity to cure the patient. Rather, withholding therapy preserves the opportunity to use it effectively in the future to control the tumor. Fitting the data for a patient to a model of cancer's evolution allows clinicians to forecast the likely results of different clinical decisions (with estimates of uncertainty)[77]. It is helpful to fit models of sensitive and resistant cell dynamics to tumor burden measures in order to forecast the expected results of continuing the current treatment versus changing treatment[77–79].

*Prefer non-chemical control methods*

Pest managers divide these different forms of control into the habitat, biological and mechanical controls. To the extent that we can control a tumor without chemicals, we can avoid selecting for therapeutic resistance.

*Habitat controls*

Habitat controls aim to change the habitat to make it unfavorable for the pests, and as favorable for the crops as possible. In IPM these are called "Cultural Controls", such as sanitizing soil, rotating crops, and choosing the best harvesting and planting time. In addition, pest managers use refuge crops to prevent or delay pest resistance[80]. Translated to oncology, this would mean making the microenvironment as unfavorable to the cancer cells and favorable to the normal cells as possible. In some cases, we might be able to exploit the loss of functions common in cancers. Fasting may cause normal cells but not cancer cells to become quiescent, thus protecting normal cells from cell cycle-specific drugs[81]. Similarly, timing therapy so that treatment is delivered when the circadian clocks of normal cells are in a more quiescent state can also serve to enhance therapeutic efficacy when cancer cells have lost their circadian rhythm[82,83]. Anti-angiogenic drugs are designed to make the microenvironment less hospitable for cancers, though they have had mixed success[84]. Hyperthermia therapy is also designed to target the cancer cell environment[85,86]. Suppression of cancers might also be accomplished by modifying the pH of the tumor[87], limiting critical resources that the cancer cells need[88,89], and other strategies for suppression such as an amino acid depletion[90].

### *Mechanical controls*

In pest management, mechanical controls include physically removing pests, constructing physical barriers to pest invasion, and using pheromone traps[91]. Surgery is a common technique for physically removing cancer cells[92]. Removing the primary tumor and metastases reduces the size of the cancer cell population and thereby reduces the evolvability of cancer. This should be beneficial even if not all metastases can be removed. However, the benefits must be weighed against the morbidity and risks of the surgery. Cancer traps, potentially including chemoattractants for cancer cells, might complement other control mechanisms and help prevent further metastases[93,94].

### *Biological controls*

Biological controls in IPM are based on the ecology of the pest, and typically include the use of natural predators, parasites, and pathogens of the pest[60,95]. Biological control in cancer treatment includes immunotherapy (predators) as well as viral and bacterial therapy. Engaging predation from the immune system may include cancer vaccines[96], immune checkpoint modulators[97], and CAR T-cell therapy[98,99]. Oncolytic viruses, by infecting and replicating in the tumor cells, can cause a reduction in the tumor without harming normal tissues, for example by producing viruses that only replicate in tumor cells with up-regulation of the RAS pathway[100] or inactive TP53[101]. As an additional benefit, a viral infection of cancer cells can be detected by the immune system, triggering an adaptive immune response in addition to an innate immune response[100]. Anaerobic bacteria are also intriguing potential anti-cancer agents that target the hypoxic parts of the tumor that are the hardest to treat with chemicals and radiation[102,103]. While any one of these forms of therapy may not be adequate to control a tumor, the mechanism of resistance to a biological control[104–107] is likely to be very different from the mechanism of resistance to chemical control, and thus they are likely to work well together.

The important factor is both biological and chemical controls should work compatibly with each other. They should be "integrated" as in the progenitor concept of IPM, "Integrated Control" proposed by Stern et al. 1959[16]. Therefore, special care should be taken when choosing a chemical control, as in agricultural IPM, such that the chemical control does not harm natural enemies or otherwise diminish biological control function. Today, as in medicine, we have very targeted pesticides that can be very specific to one or a few pests.

### *Use drugs as specifically as possible*

Using drugs specific to a pest or cancer helps to minimize off-target effects. For example, insect pheromones (e.g., sex pheromones) that are specific to the pest and interfere with pest mating have less toxicity for non-target species than other exogenous chemical treatments[60]. In addition, there are growth, development, and behavior-modifying insecticides, like "insect growth regulators" and feeding inhibitors, and occasionally repellents, and more commonly the specificity of the agent is such that its effects are a very narrow spectrum [108]. Because cancer cells are derived from human cells, it is inherently difficult to develop cancer-specific therapies that do not affect normal cells. However, there have been a number of successful therapies that target either the cell type[109], a driving gene[110–112], or a carcinogenic mutation[113,114]. Certain behavioral aspects of cancer cells may also be targeted. Selectively inhibiting matrix metalloproteinases prevents the destruction of the extracellular matrix (including basement membranes) and thereby helps prevent invasion[115]. By harnessing this phenomenon, keeping a tumor localized may enable treatment through surgery.

*Use the lowest dose possible*

The principle developed within the IPM framework is to use the minimum effective dose - the lowest dose that can still control the pest population. Limiting chemical use is accomplished by 1) limiting the frequency of use or the number of sprays, because each is a selection force on the pest population, and 2) limiting the dosage to the lowest effective dose. This too limits selection pressures, saves money, and likely reduces side effects. This should be applicable to cancer therapy by applying only enough drugs to prevent the tumor from growing but no more.

Most chemotherapeutic drugs can be characterized by a dose-response curve: the higher the drug dose delivered, the more cancer cells it kills[116,117]. This relationship between dose and cell death is limited at high doses by diminishing returns as well as unacceptable toxicity[116,117]. This relationship naturally led to the use of the maximum tolerated dose in most cancer therapy protocols[118,119]. Unfortunately, we know from evolutionary theory that the maximum tolerated dose is also the fastest way to select for acquired therapeutic resistance[120]; the strength of selection is directly related to the speed of evolution. This observation implies that we can slow the rate at which therapeutic resistance evolves by lowering the dose of the drug.

*Reduce cross-resistance*

*Diversify the use of MoAs as much as possible*

In the IPM strategy pest managers use diversified MoA wherever possible such that they are not over-reliant on a single or small set of modes of action. Cancer drugs may be grouped by the MoA they use to achieve their anticancer effects. Both theory and experiment show that resistance to one drug with a particular MoA also confers cross-resistance to other drugs that use the same MoA[121,122]. Thus, to avoid cross-resistance and reap the benefits of using multiple drugs, those drugs should be chosen so that they use different MoAs[22]. This principle has been recognized in oncology, though it is not always followed[123,124]. In oncology, there are a variety of agents with different MoAs including (1) antimetabolites (e.g., methotrexate), which interfere with DNA synthesis by mimicking nucleotides; (2) alkylating agents (e.g., carboplatin) that bind to DNA causing DNA breaks, triggering apoptosis; (3) antimicrotubule agents (e.g., docetaxel) that prevent cell division by interfering with microtubules; (4) topoisomerase inhibitors (e.g., doxorubicin) that interfere with topoisomerases, causing DNA breaks or inhibition of DNA synthesis; and (5) anti-angiogenic agents (e.g., bevacizumab) which prevent tumor growth by inhibiting the formation of new blood vessels that tumors require to absorb nutrition and oxygen for growth. Radiotherapy represents an entirely different MoA that also kills by damaging DNA (through small breaks in the DNA inside cells) as well as radiolysis of water, and may enhance an immune response after exposure to radiation[125–128]. We might be able to do even better in improving patients' survival by combining drugs for which resistance to one drug causes sensitivity to the other, and vice versa, an approach called double-bind therapy[22,61,129,130].

*Partition MoAs in space or time so as to segregate their usage as much as practically possible*

The combination of drugs with different MoAs does not solve the problem of acquired therapeutic resistance. The hoped-for efficacy of multidrug therapy is predicated on the assumption that different mutations are necessary to generate resistance to different MoAs. Unfortunately, we have known for a long time that single alterations can cause multidrug resistance[131] in the form of upregulated efflux pumps that can expel a variety of drugs from the cell before their various MoAs can cause cell death[131]. Combinations of pesticides select for similar multi-mode of action resistance mechanisms in pests[132,133], so pest managers advise separating the application of different drugs as much as possible in both time and space. This suggests that we should use drugs with rapid half-lives so that one drug is completely cleared from the system before the cancer is exposed to the next drug[134].

Perhaps the most striking difference between oncology and pest management is that pest managers try to never apply the same MoA twice in a row, and never more than twice within a growing season (e.g., the Arizona Cotton IPM strategy) (Figure 2). Overall, in the IPM industry, pest managers endeavor to never apply the same MoA twice in a row to the same generation of insects, and different MoA is used wherever possible. In oncology, we repeatedly apply the same drug or drug combinations, week after week. In contrast, pest managers endeavor to switch pesticides every time they spray a field[11]. This prevents a resistant clone from expanding much before it is exposed to a new drug with a different MoA.

It is unclear how to partition modes of action by space in cancer. In agriculture, different fields can be sprayed with different pesticides. In addition, precision agriculture is developed, so even within a field, areas of a field can be targeted precisely with a pesticide "spot" spraying. therefore, by only treating portions of the field where the pest is at unacceptable levels and leaving the rest of the field unsprayed and therefore unselected for resistance. In oncology, it is more difficult to apply and contain an intervention to a particular location, with the exception of radiation therapy. However, different drug delivery systems (such as nano-materials and exosomes[135]) or the use of anaerobic bacteria that only thrive in hypoxic areas[136] may lead to spatial partitioning of drug exposures.

[FIGURE 2]

Figure 2: Schematic of a MoA schedule in pest management. This schematic illustrates principles of integrated pest management. Pest managers avoid sequential use of the same MoA, and to avoid a MoA more than twice within the same growing season. For example, in the figure, drugs with MoA1 and MoA2 are reused in the same growing season but only after other MoAs have been used. To avoid selecting for acquired therapeutic resistance in cancer, we should separate repeated applications of a MoA as much as possible.

Pest managers may mandate that a portion of the fields not be sprayed at all so as to preserve sensitive pests. This can be an effective strategy for controlling resistant pests because resistance typically comes with a fitness cost in the absence of the pesticide[61,137]. Preserving sensitive pests allows them to out-compete resistant pests, which preserves the ability to control the pest population in general. This is a combination of the principles of IPM: (1) not treating if the pest burden is acceptable, (2) using the minimum effective dose when you need to use a chemical control (limiting in both frequency and dose), and (3) segregating the drugs in space. This idea of preserving sensitive pests inspired Gatenby to develop adaptive therapy for cancers (see below).

*Evaluate success based on long-term control*

Prior to the development of IPM, farmers would generally evaluate pest management methods by a combination of crop yields and the absence of pests[11]. Practices prior to IPM were usually calendar-based, so-called, "womb-to-tomb" spraying, in which pesticides were used until the end of the production cycle [138]. Successful management of resistance in pests required a shift in evaluation criteria away from short-term response and the impractical goal of total pest eradication to the more helpful goals of long-term management, cumulative yield, and stability of yield over time[11]. The parallel for oncologists may be to move away from partial and complete response as the criteria of evaluation and focus on time to progression, survival time, and quality of life.

*Forecasting*

Forecasting is central to IPM efforts and it informs much of what pest managers do including decision-making, monitoring, prevention efforts, managing resistance, and evaluation. However, formal forecasting is rarely applied in oncology. Gu et al., (2012) studied patients with glioblastoma multiform and demonstrated the impact of a predictive model based on patient-specific planning of the treatment strategy[139]. It is helpful to fit models of sensitive and resistant cell dynamics to tumor burden measures in order to forecast the expected results of continuing the current treatment versus changing treatment[77–79]. The more frequently the tumor can be monitored, the better those forecasts can be[140].

**The spread of resistance management in oncology**

Best practices in resistance management have not been universally adopted in pest management systems. In some cases, local successes have spurred the spread of these techniques. In other cases, a top-down approach of regulations banning counter-productive approaches and enforcing good pest management principles has been necessary to achieve collective success.

In IPM, we have learned that only using the perceived "best" drug each time will rapidly lead to its ineffectiveness because of resistance. Sometimes "best" is merely perception and there are equally effective but different treatments available. Other times, we have to encourage switching from the best treatment to the second best one just to limit the evolution of resistance.

But in most cases, effective pest management has often spread through the desperation engendered by disasters[141]. We hope that the success of well-designed clinical trials in extending survival time and quality of life will drive the adoption of pest management principles in oncology through evidence-based medicine, but a change in practice from trying to shrink tumor burden to stabilizing tumor burden and living with it is likely to face significant emotional and psychological hurdles in both patients and physicians.

*Initial efforts to translate pest management to oncology: Adaptive Therapy*

Clinical trials of IPM principles have only just begun in oncology. To date, only one has been completed: a pilot clinical trial of adaptive therapy in castration-resistant, metastatic prostate cancer[39,142]. Adaptive therapy seeks to prevent the expansion of therapeutically resistant clones by maintaining chemosensitive cells in the tumors to compete with the resistant cells[37,38,120,143]. This is done operationally by trying to keep the tumor at a stable size. In the dose modulation version of adaptive therapy, the tumor burden is measured frequently and the dose is raised if the tumor is growing, but the dose is lowered if it is shrinking. Otherwise, if the tumor is stable, dosing is kept at the same level (Figure 3). This protocol was able to keep tumors stable indefinitely in xenograft mouse models of ovarian cancer[143], triple-negative breast cancer[37], and ER+ breast cancer[37]. Furthermore, the dose necessary to control the tumor decreased over time. This might be explained by the fact that the tumors in the adaptive therapy arms evolved better perfusion than the tumors in the maximum tolerated dose (MTD) arms[37]. The clinical trial of adaptive therapy used a different protocol, similar to intermittent therapy, in which abiraterone was given at MTD until the tumor burden (measured by prostate-specific antigen (PSA) in the blood) fell below 50% of the initial tumor burden. At that point, the dosing was stopped until the tumor burden grew back to 100% of its initial burden, at which point dosing was started at MTD again, and so on[39]. This resulted in a doubling of radiographic progression-free survival[142].

**[FIGURE 3]**

**Figure 3. Comparison of standard therapy with two models of adaptive therapy.** In standard therapy (A), doctors use the maximum tolerated dose (MTD) of a drug. In dose modulation adaptive therapy (B) the dose of the drugs is adjusted based on the tumor's response. In another form of adaptive therapy, intermittent(C) (treatment skipping), the MTD of the drug is used until the tumor shrinks, then the treatment is skipped.

*An illustrative example: an IPM-inspired clinical trial of colorectal cancer*

The principles and approaches defined in IPM may be extrapolated to any type of cancer. Here we describe how these principles might be employed in the design of a clinical trial for colorectal cancer. We chose colorectal cancer as an example because 1) it is common[144], 2) It is difficult to manage the late stages of this disease[145], and 3) there are a variety of drugs that can produce an initial response[146], and so might be rotated, or adjusted in an IPM-inspired protocol. Inclusion criteria for a therapeutic protocol should prioritize stage III and IV patients that have not received prior chemotherapeutic interventions, or for which only one MoA has failed and where complete resection is either not possible or was unsuccessful.

1. *Decision-making (consider the patient's biology, drug, and target cells)*

    An ideal trial based on IPM should start with profiling the patient's cancer to determine what known forms of resistance are present in the tumor as well as what targetable mutations are present and at what frequencies. This might involve extensive multi-region sequencing, or deep sequencing of ctDNA if ctDNA provides a reasonable measure of the tumor cell population. Such profiling should not rely on a single biopsy but multiple biopsies, as a single biopsy is unlikely to be representative of the entire cancer cell population[147].

2. *Use drugs as specifically as possible*

    If sequencing or other profiling of cancer identifies promising drugs specific to clonal targets in cancer, they should be included in the rotations, along with the more broad-spectrum therapies. A subset of colorectal cancers responds well to epithelial growth factor receptor (EGFR) inhibitors. Targeted drugs typically have less toxicity than chemotherapies but inevitably lead to the evolution of resistance. This makes them ideal for IPM approaches which are designed to prevent therapeutic resistance but might require the long-term application of low doses of drugs.

3. *Choose drugs so as to reduce cross-resistance*

    We identified existing FDA-approved drugs for colorectal cancer with four distinct MoAs: DNA damaging agents, including DNA synthesis blockades and platinum therapies such as oxaliplatin, anti-metabolites such as 5-fluorouracil and Capecitabine, epithelial growth factor receptor (EGFR) inhibitors such as Cetuximab and Panitumumab, and vascular endothelial growth factor (VEGF) inhibitors such as regorafenib and bevacizumab. Further, MoAs may be added if any additional targets can be identified in a particular cancer. A cycle would include the application of one drug at a time, applied once, and then followed by a one-week recovery time before the next drug with a different MoA is applied. After 3 weeks (with one drug each week), the patients would receive a further week break, to complete the 4-week cycle. The third week in each cycle would be an anti-angiogenic drug (which restricts the tumor vasculature) (Figure 4).

The use of an anti-angiogenic drug on the third week of each cycle is inspired by the mechanical controls used in IPM. Restricting the vasculature of the tumor after the other two drugs in a cycle ensures that the earlier drugs can be delivered to the tumor. Then, by restricting the tumor vasculature we seek to create a resource-depleted environment. This restriction in resources makes it especially hard for resistant cells to grow in comparison to sensitive cells since the former often requires more resources to pay the cost of resistance[37].

**[FIGURE 4]**

**Figure 4**. Proposed scheduling of drugs that applies each drug once per 4-week cycle, and uses different MoAs. The third week of each cycle uses an anti-angiogenic drug, so as not to restrict the vasculature before trying to deliver the other drugs. Week 4 of each cycle is a drug holiday break to allow the patient to recover.

4. *Use the lowest dose possible*

Phase I trials, de-escalation trials, and metronomic therapy trials for each drug should be re-analyzed to identify the minimum dose that reliably results in a measurable response or stable disease. This should be the starting dose for each drug but should be adjusted during treatment as necessary to maintain a stable disease. Furthermore, a lower threshold should be established, such that if the tumor burden falls below that threshold, treatment will be halted and the patient monitored until the tumor burden grows above that threshold again. Ultimately, this result leads to limiting drug use to both the lowest effective dosages and the fewest number of times to accomplish the goal. Simulation studies suggest that the higher that threshold (*i.e.*, the sooner we pause treatment) the better[148].

5. *Prefer non-chemical control methods*

Treatment should start with surgery where feasible. To the extent that the primary tumor and any metastases can be safely removed, they should be (a mechanical control). This reduces the population of cancer cells that can evolve, and so may both remove drug-resistant cells and reduce the probability that new mechanisms of resistance will evolve. Ultimately, however, those benefits must be weighed against the toxicity and morbidity of the surgery. To manage the adverse effects of cancer treatment on patients, improve quality of life, and potentially shift the tumor microenvironment to benefit the normal cells, appropriate supportive care intervention might be used including physical exercise, diet and nutrition, and psychological counseling [149].

6. *Monitor continuously*

It is unclear how often a tumor should be monitored during treatment as this will vary between patients and tumor types. It may even vary temporally within a given patient as the dynamics of their cancer slow down or speed up. What we know from IPM is that the disease should be monitored frequently and interventions should be adjusted in response to what is evolving in cancer. That frequency will likely be limited by the expense of the tumor monitoring method, the willingness and ability of patients to attend clinic visits, and any toxicity of the assessment (e.g., radiation from CT scans[150]). Technological advances may well alleviate these concerns with minimally invasive assays of tumor burden and wearable or home-use monitors.

Drugs that do not reduce or maintain stable tumor burden would be removed from the treatment rotation. If the tumor burden continues to grow during the application of a particular mode of action, drugs with that mode of action may be removed from the rotation. This might require monitoring at the same frequency as drug switching. In some cases, it might be difficult to determine whether the tumor burden is increasing, due to the potential for immune infiltrates, fibroblasts, extracellular matrix, and necrosis to increase the size of a tumor without increasing the number of cancer cells. Ideally, measures of the number of cancer cells should be used, perhaps with tumor-specific blood markers.

Monitoring, and treating tumors with a new drug only when the tumor burden is above the treatment threshold should continue indefinitely to control tumor growth for as long as possible. The primary outcome for this study should be overall survival, with secondary outcomes including measures of quality of life and time to progression (defined as the tumor continuing to grow despite the use of all available drugs). The goal here is to manage cancer as a chronic disease so that patients can live with their cancers but not die from them, or from the treatments.

*Summary of clinical trial*

Pre-clinical trials using mouse models would inform us about the effectiveness of our protocol, though those mouse models must demonstrate acquired therapeutic resistance under the standard of care in order to faithfully represent the clinical challenge. Colorectal cancer remains an appealing candidate disease due to the existence of well-described spontaneous murine models[151]. If preclinical experiments with the above design are successful, clinical trials may be able to bypass phase 1 due to the fact that safe dosage ranges for the drugs in our design would have already been determined. For phases 2 and 3, patient recruitment should involve those in the second line of treatment (and exclude drugs with a similar MoA from the first line of treatment that was used on those patients).

*Challenges and clinical needs for IPM-inspired oncology*

In order to develop IPM-inspired treatments for cancer, there are a few important challenges that will need to be addressed. First, most IPM approaches will require frequent and accurate measurements of tumor burden. It is not clear how frequently we will need to measure that tumor burden to appropriately adjust our interventions. The fact that adaptive therapy and other IPM-inspired therapies require regular tumor burden measures also means that we will need to find economical and safe ways to measure it. Second, we need biomarkers to help predict if a cancer is curable or if we should focus on management instead (and hopefully what types of control should be tried). In particular, if we could distinguish cancers that are likely to harbor therapeutically resistant clones from cancers that do not, we could focus on controlling the former and curing the latter.

*Measuring tumor burden (frequently)*

There are several radiological and non-radiological methods for assessing tumor burden. Radiological methods such as X-rays, CT scans, PET scans, MRI, and ultrasound all have different strengths and weaknesses[152]. When used in conjunction with RECIST 1.1 criteria, they only measure one dimension of the lesions, not even tumor volume[70]. They also typically measure only a few but not all lesions in a patient[70]. There are tools for radiometrically measuring tumor volume[153,154], but those measurements may not be accurate for a variety of reasons including subjective assessment of tumor boundaries, necrotic regions within the tumor, and non-cancer cells such as lymphocytes, fibroblasts, adipocytes, etc. that all contribute to tumor volume. Furthermore, these measurements are typically made months after the completion of a therapeutic protocol, not during treatment, which makes it impossible to change interventions during treatment, as is standard for IPM. If we are to measure tumor burden frequently, issues of cost, convenience, and toxicity of the measurement all become practical challenges to the deployment of IPM-inspired therapies.

Additionally, the measurements of tumor burden in solid tumors and liquid tumors are different. In acute myeloid leukemia (AML), measures of cancer cells in the blood are sufficient to monitor tumor burden[155]. In multiple myeloma, assessment of the M-spike protein in the blood is used for tumor monitoring[156]. Several tumor markers (cancer biomarkers) in the blood are used to assess cancer growth in solid tumors[157–159] like PSA, which is standard in prostate cancer and was used in the prostate cancer adaptive therapy trial[39].

Circulating tumor DNA (ctDNA) in the blood is a promising, minimally invasive biomarker for monitoring cancer[160]. It may not only measure tumor burden but also be more generally representative of the cancer cell population and molecular diagnosis (mutations of the tumor)
then a single biopsy[160]. It may also reveal mutations that cause resistance and thus enable monitoring of some resistant populations as well as overall tumor burden[74,161,162]. Circulating tumor cells (CTCs) may also be effective for monitoring late-state disease, and resistant clones, and identifying the mechanisms of resistance[163–165].

*Biomarkers for cure versus control: Detecting resistant clones*

In IPM, when pest managers identified a resistance potential or gene, there are precious few programs where monitoring is contemporaneous with management. In pest management, they do posthoc, post-season reviews and adjust "therapies" in the next season. Therefore, they focus on this principle, limiting the use of the pesticide to the lowest levels possible (that could be frequency or dosage or both, but is itself central to IPM). And, once chemical controls are needed, they advocate for diversified use of modes of action. Moreover, where possible, pest managers seek to manage refugia through space and time, which requires partitioning chemistry throughout the season or across the community. Finally, even without perfect information about the status of any given resistant clone, we assume that resistance is constantly a threat and constantly evolving and therefore responsive to the first principles mentioned above. I.e., we endeavor to always be practicing IPM and integrated resistance management.

Ideally, we would develop biomarkers that would help determine if a cancer is likely to be curable or, alternatively, will likely require management strategies to control the disease. If clones resistant to a particular drug are already present in a neoplasm, treatment with the intent to cure with that drug is doomed to failure and efforts should shift to alternative drugs and/or control of the tumor. Unfortunately, it is fundamentally difficult to detect if a resistant clone is present. Resistant clones may be so rare in a neoplasm (e.g., 1 in $10^9$ cells) as to be below the level of detection for most technologies. Furthermore, there are often multiple mechanisms of resistance for any given drug[131,166] and it is likely that many mechanisms of resistance remain to be discovered, making it hard to recognize resistant clones.

Although developing direct methods for detecting resistance is difficult, there are a number of indirect methods that may be sufficient to guide patient care. It is likely that a highly diverse cancer will harbor a resistant clone and indeed measures of intratumor heterogeneity (ITH) predict survival in many cancers[2,167,168]. One advantage of using measures of ITH to predict the presence of a resistant clone[169]. It is also possible to infer the presence and kinetics of therapeutically resistant clones through longitudinal tumor burden data. When cancer's response to therapy diminishes, that is an indication that the proportion of sensitivity to resistant cells is decreasing. By fitting the longitudinal tumor burden data to simple models of sensitive and resistant cell dynamics, we can estimate the resistant clone's size and growth rate[77,170]. Some forms of resistance require a dramatic increase in glucose metabolism that can be detected on PET scans[159,171,172]. Finally, the growth of a resistant clone might be detected via passenger mutations in that clone appearing in ctDNA, even if the ctDNA does not reveal the mechanism of resistance[72–76].

**Conclusions**

Drug resistance is one of the most important problems we face in clinical oncology. Pest managers face the same problem with pesticides selected for resistant pests. In both cases, a single mode of action therapy tends to fail. Multi-mode of action therapies are more effective than single-mode of action therapies but still lead to multidrug resistance after a short period of time. However, integrated pest management dynamically slows the development of pesticide-resistant pests often below densities requiring control over the long term. Non-chemical and chemical approaches used in IPM may be translated to cancer therapy. Modern principles of resistance management could control the growth of resistance by limiting the use of each mode of action (MoA) (minimizing the frequency of use and dose to the lowest practical levels), diversifying MoA use (through rotation programs), partitioning MoA use in space or time, and utilizing non-chemical methods of control.

In fact, a general principle underlying most of the lessons of IPM is that the more drugs we use, the more we select for resistance. Most strategies of IPM are essentially ways to limit the amount of drug used, while still controlling the infestation. Integrated pest management inspired adaptive therapy in oncology[37,120,143]. An ideal trial based on IPM would start with intensively monitoring cancer and rotating drugs at minimum effect doses, avoiding using any one MoA for so long as to select for resistance. If we assume that resistant cells are already present, as pest managers do, this approach represents a fundamental shift from trying to cure cancer to managing it, with the aim of prolonging patient survival and quality of life. By transcending the sole focus on a cure, we open up new possibilities for methods to dramatically improve both patient survival and quality of life.

## Acknowledgments

Thanks to Gabrielle Hirnese for help with the figures. This work was supported in part by NIH grants U54 CA217376, U2C CA233254, P01 CA91955, and R01 CA140657 as well as CDMRP Breast Cancer Research Program Award BC132057 and the Arizona Biomedical Research Commission grant ADHS18-198847. The findings, opinions, and recommendations expressed here are those of the authors and not necessarily those of the universities where the research was performed or the National Institutes of Health. This material is based upon work that is supported by the National Institute of Food and Agriculture, U.S. Department of Agriculture, under award number 2021-70006-35385

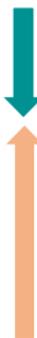

```
Lessons from IPM
  │
  ▼
  ↑
To Oncology
```

- **Prevention and suppression**
  - Cancer prevention

- **Decision making**
  - Consider the patient's biology, drug and targeted cells
  - Do not treat small or benign tumors

- **Monitoring the problem**
  - Frequent evaluation of tumor scans and biomarkers

- **Reduce cross-resistance**
  - Use different MoAs, and include biological controls

- **Select pesticides with least side effects**
  - Prefer low toxicity drugs and drugs without cumulative toxicity

- **Reduce pesticide use**
  - Use minimum effective dose

- **Prefer non-chemical control methods**
  - Habitat controls: e.g., hyperthermia therapy
  - Mechanical: e.g., surgery
  - Biological: e.g., CAR-T cell therapy

- **Evaluation of treatment**
  - Evaluate success based on long-term control

- **Forecasting**
  - Forecast tumor response based on measurements to date

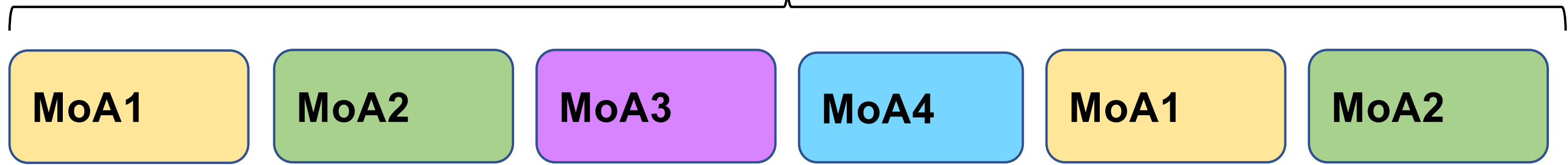

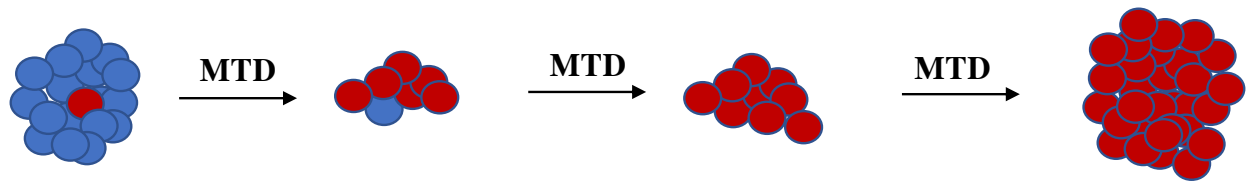

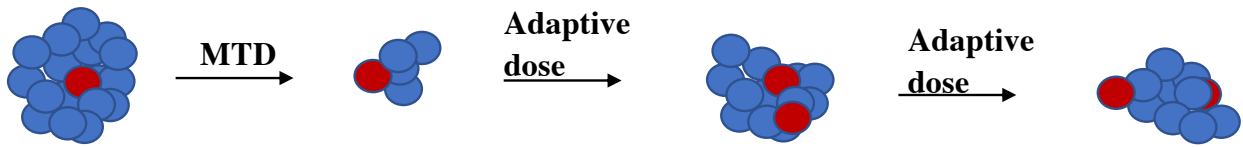

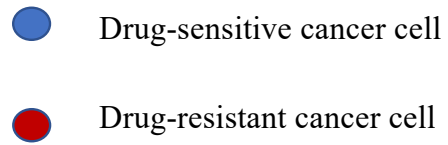

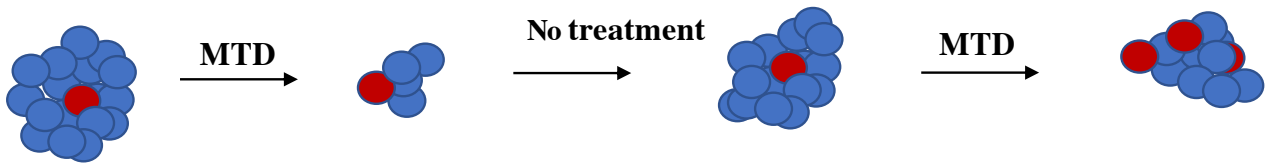

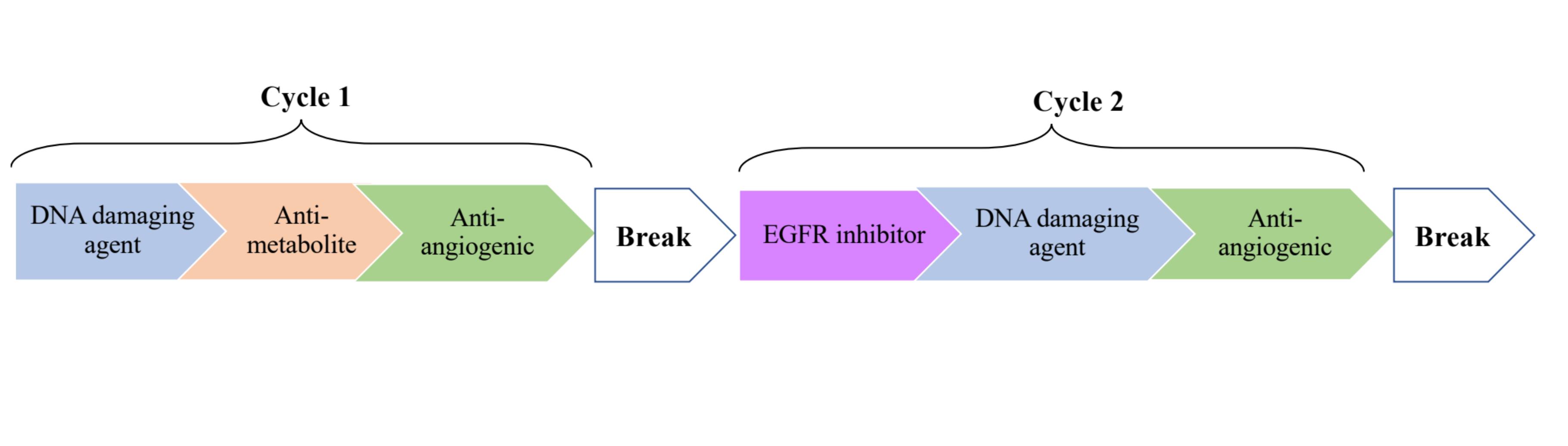